\documentclass[sigconf]{acmart}

\usepackage{array,graphicx}
\usepackage{float}
\usepackage{multirow}

\AtBeginDocument{%
  \providecommand\BibTeX{{%
    \normalfont B\kern-0.5em{\scshape i\kern-0.25em b}\kern-0.8em\TeX}}}

\newcommand{\jieshan}[1]{\textcolor{teal}{#1}}

\setcopyright{acmcopyright}

\acmConference[CHI EA '24]{Make}{May 11–16, 2024}{Honolulu, HI, USA}
\acmBooktitle{Extended Abstracts of the CHI Conference on Human Factors in Computing Systems (CHI EA ’24)} 
\acmDOI{}
\acmISBN{}

\begin{document}

\title{MAxPrototyper: A Multi-Agent Generation System for Interactive User Interface Prototyping}

\author{Mingyue Yuan}
\affiliation{%
\institution{University of New South Wales}
  \country{Australia}
}
\email{mingyue.yuan@unsw.edu.au}
  
\author{Jieshan Chen}
\affiliation{%
  \institution{CSIRO’s Data61}
  \country{Australia}
}
\email{Jieshan.Chen@data61.csiro.au}

\author{Aaron Quigley}
\affiliation{%
  \institution{CSIRO’s Data61}
  \country{Australia}
}
 \email{Aaron.Quigley@data61.csiro.au}

\begin{abstract}

In automated user interactive design, designers face key challenges, including accurate representation of user intent, crafting high-quality components, and ensuring both aesthetic and semantic consistency. 
Addressing these challenges, we introduce MAxPrototyper, our human-centered, multi-agent system for interactive design generation. 
The core of MAxPrototyper is a theme design agent. It coordinates with specialized sub-agents, each responsible for generating specific parts of the design.
Through an intuitive online interface, users can control the design process by providing text descriptions and layout. 
Enhanced by improved language and image generation models, MAxPrototyper generates each component with careful detail and contextual understanding. 
Its multi-agent architecture enables a multi-round interaction capability between the system and users, facilitating precise and customized design adjustments throughout the creation process.

\end{abstract}

\ccsdesc[500]{Human-centered computing~Systems and tools for Interaction design; Computing methodologies~Intelligent agents}
 
\keywords{Interactive design, user interface, conversational agents, large language models, image generation}


\maketitle



\section{Introduction}

User interface (UI) design is an essential aspect of the modern software industry, as it plays a significant role in shaping the overall user experience. While traditional UI design approaches require extensive effort and expertise, numerous studies~\cite{chen2020wireframe, zhao2021guigan, 2021OwlEyes, 2021Screen2Vec, swearngin2019modeling} have explored and validated that the use of techniques, such as deep learning, can alleviate the workload associated with UI design. Despite these advancements, there is still a disconnect between the tools available and the daily practice of designers.

UI design today has advanced significantly, for example with the use of low-fidelity visual models such as wireframes to help create high-fidelity functional prototypes~\cite{chen2020wireframe}. In addition, techniques such as Swire~\cite{2019Swire}, Screen2Vec~\cite{2021Screen2Vec}, and VINS~\cite{2021VINS} incorporate sketches or screenshots to retrieve GUIs, while Guigle~\cite{2019Guigle} and RaWi~\cite{2023Kolthoff} further enhance retrievability, thus simplifying the prototyping process. However, these retrieved GUI images often suffer from limitations in terms of visual fidelity, creativity and reusability.

Widely used GUI prototyping tools, such as Sketch~\cite{sketch}, Adobe XD~\cite{adobe}, Figma~\cite{figma}, typically offer a combination of fundamental GUI components and templates. However, their inability to generate customized results also affects their support for the creative process and efficiency of the overall design process. Additionally, while generative methods such as layout2image~\cite{biswas2021docsynth}, VAE~\cite{razavi2019generating}, MidJourney~\cite{midjourney} and stable diffusion~\cite{rombach2022high} showcase remarkable creative potential, their practical implementation is hindered by difficult to control. resulting in unstructured and non-editable outcomes. The need for manual reconstruction and the vagueness of components occupying small areas within the overall image generation, such as ``text'', ``text button'', and ``icon'', further complicate these issues.

To address the limitations of non-editable outputs, enhance the creative potential of retrieval-based methods, and guarantee high-quality prototype generation,
we present our novel interactive design system \textbf{MAxPrototyper} — Multi-Agent Collaboration for Explainable UI Prototype Generation. This system empowers designers to craft high-fidelity prototypes that are both customizable and comprehensible at each stage of the generation process.

MAxPrototyper initiates the design journey with user-provided inputs such as text descriptions and a wireframe layout. 
The text provides the general design requirement (e.g., ``Starting page for MAxPrototyper: a intelligent design assistant.''), and the wireframe gives a preliminary layout of the intended UI design.

Specifically, this process draws inspiration from a top-down design philosophy. It allows our system, MAxPrototyper, to first generate an overall theme design, followed by the iterative creation of each individual components. This process ensures aesthetic consistency throughout the design generation, thus maintaining a coherent visual style across all elements of the prototype.

A core feature of our system is the ability of a central agent to utilize a cache pool, which blends accumulated results to effectively guide the sub-agents' actions, thus maintaining the integrity of the design context. This strategy enables accurate multi-round interaction capabilities. Designers are thus provided with an iterative framework that supports an in-depth understanding and customizable adjustments at each generation phase.

By balancing automated generation and customization, we aim to support a streamlined design process and tool that empowers designers with deeper insights and greater control. The resulting prototypes can be conveniently saved in SVG or JSON format, further empowering designers with practical efficiency.

\section{Method}
\label{sec:approach}

\vspace{-3mm}
\begin{figure}[h]
  \centering
  \includegraphics[width=0.48\textwidth]{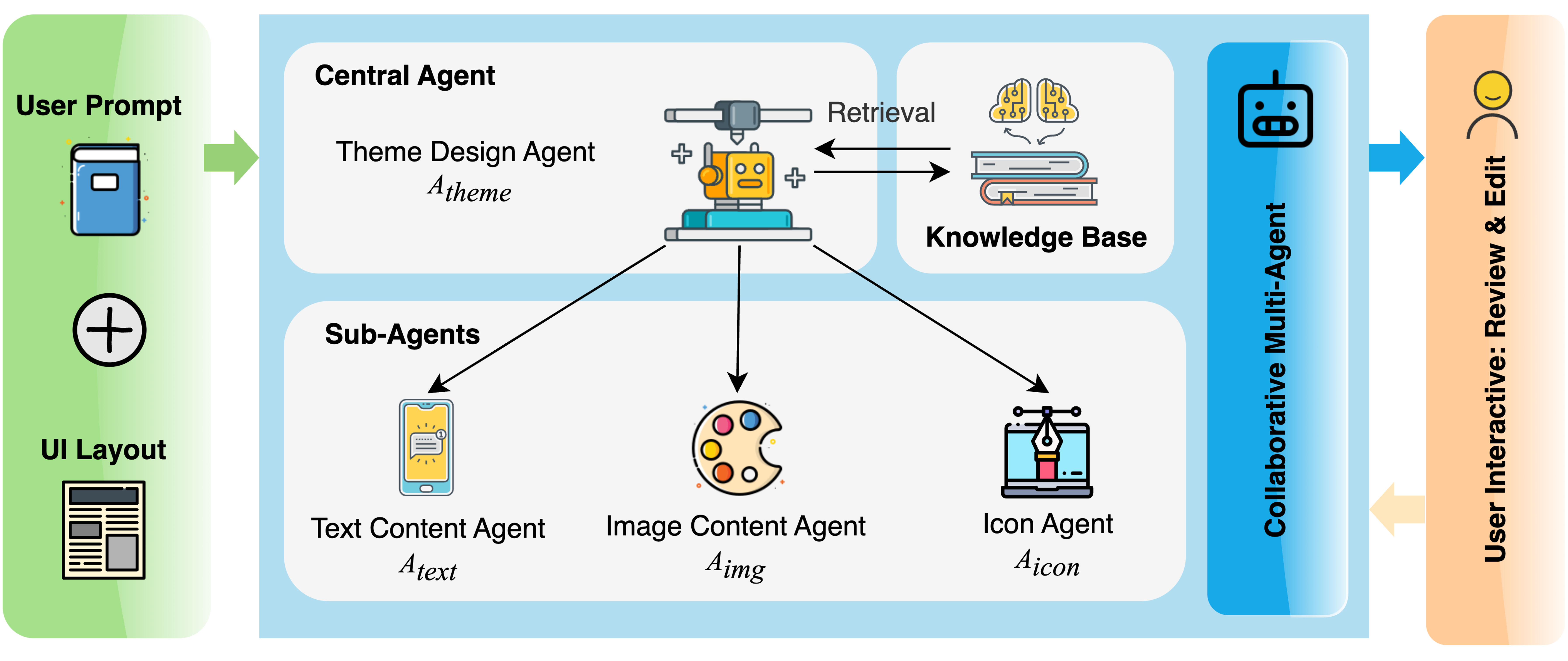}
  \caption{ Overview: using user prompt and UI Layout as input, MAxPrototyper utilizes a multi-agent method for UI prototype creation. It integrates four primary agents—Theme Design, Text Content, Image Content, and Icon—with Theme Design as the central guiding the sub-agents.
  }
  \label{fig:overview-MAx}
\end{figure}
\vspace{-3mm}

As illustrated in Fig.~\ref{fig:overview-MAx}, our system consists of knowledge bases and four key agents: Theme Design Agent \( A_{theme} \), Text Content Agent \( A_{text} \), Image Content Agent \( A_{img} \) and Icon Agent \( A_{icon} \). 
We collect knowledge bases to import relevant design knowledge.
Building upon this foundation, MAxPrototyper processes user inputs, comprising both a prompt and a UI layout. The system initiates its workflow with the Theme Design Agent, which formulates a high-level theme description and generates a corresponding theme image to establish a coherent and consistent design narrative.

Served as the main director, \( A_{theme} \) subsequently execute specialized tasks through \( A_{text} \), \( A_{img} \) and \( A_{icon} \) for the detailed creation of each component. Seamless coordination between the main agent and supporting agents is vital for effectiveness. Utilizing a blend of previous results and a cache pool, the central agent effectively informs the sub-agents, ensuring the design context is maintained.

\subsection{Knowledge Base Construction}
\label{sec:knowledge}
Employing domain-specific knowledge enhances the creativity and output quality of Large Language Models (LLMs).~\cite{wang2023voyager, lu2023chameleon}. 
We collect two knowledge bases, namely UI knowledge (pairs of theme descriptions with local component descriptions) and Icon Knowledge (Pairs of SVG code with descriptions) for \( A_{theme} \) and \( A_{icon} \).

\subsubsection{\textbf{UI Knowledge Base}} 
\label{sec:uiknowledge} 
\textit{}

\textit{\textbf{1) UI Composition Knowledge.}}
Rico Dataset~\cite{rico} is one of the most comprehensive open-source UI datasets, 
including the UI screenshots, their view hierarchy information and their attributes like text, bounds and class, and the composition of these UI elements.
We extracted class and bounds from these metadata, and form the UI composition knowledge (\textit{<component types><bounding boxes>}) for each UI.

\textit{\textbf{2) UI Semantic Knowledge.}}
The fine-grained component description can be obtained by parsing content-description from the Rico dataset metadata. We obtain \textit{<text content/icon descriptions>}.

While Rico contains the metadata of the composition and text contents of UI, it lacks the high-level UI description. We obtain this data through Screen2Words~\cite{2021Screen2Words}, which augments Rico to provide ~112k high-level textual descriptions for its 22k UI screenshots. Through this dataset, we obtained \textit{<UI description>}.

Beyond UI functionality semantics, design generation necessitates thoughtful consideration of themes, colors, and the target audience. 
We adopt Blip2~\cite{li2023blip}, which is a zero-shot visual language model, to generate theme descriptions via a visual question-answering approach.
We identify four key attributes for theme design: theme color, primary color, theme description, and app category, 
and create a specific question, pairing it with a UI image, and then input it into Blip2 to derive the answer, which is shown in Table~\ref{tab:vqa}.
Finally, these three descriptions are concatenated together, and form UI Semantic Knowledge: \textit{<text content/icon descriptions><high level description><theme design description>}.

\subsubsection{\textbf{Icon Knowledge Base}}


We collected the data from Google Material Design Icons~\cite{materialicon}, a high-quality repository that stores over 900 diverse icons (in SVG format and with text description). 
We obtain our Icon Knowledge Base:  \textit{<icon SVG code><semantic description>}.



{
\vspace{-4mm} 
\begin{table}[htbp]
  \centering
  \caption{Blip2's VQA instruction templates for screenshots}
  \scriptsize   
    \begin{tabular}{ll}
    \toprule
    \textbf{Attributes} & \textbf{Blip2 Instruction Templates} \\
    \midrule
    Theme Color & \textit{``What is the background color of this screenshot?''} \\
    Primary Color &  \textit{``Besides the background, what's the dominant color in this image?''} \\
    Theme Description &  \textit{``Can you describe this screenshot in detail?''} \\
    App Category &  \textit{``Which category does this app belong to? ''} \\
    \bottomrule
    \end{tabular}%
  \label{tab:vqa}%
\end{table}%
\vspace{-5mm}
}
\subsection{Theme Design Agent}

\( A_{theme} \) acts as the supervisor, directing the UI design process. Powered with UI knowledge base collected in Section~\ref{sec:knowledge}, it sets the general style by generating the global theme description and theme image. These information can make sure the generated UI design is of high quality, coherent and consistent. 

\subsubsection{\textbf{Knowledge Retrieval}}

Domain-specific knowledge enhances the accuracy of LLM-generated content. Recognizing the LLM's token input limitations and the complexities of fine-tuning, we introduce a knowledge retrieval phase, infusing domain-specific knowledge into our generation process.

Based on the user prompt ($In_p$) and UI layout ($In_l$), we want to retrieve the most relevant knowledge from our large knowledge base.
To do so, we concatenate these two information together, and encode them into a vector $\textbf{Emb}(In)$ as the query vector, where \( In = In_p + In_l \). We use the TEXT-EMBEDDING-ADA-002 embedding model~\cite{openaiemb}).
Similarly, we also embed each piece of UI knowledge related to UI Composition and Semantic into a vector ($\textbf{Emb}(kb_j)$) as well.
Then we compute the cosine distance between query and each knowledge, and retrieve the top-k results to instruct our multi-agent system~\footnote{We set $k=2$ in our experiments.}. 
We denote the retrieved knowledge as $refer_i$. 

Our preliminary experiments, alongside findings by Bryan~\cite{wang2023enabling}, indicate that when employing related knowledge as few-shot prompting, the initial example tends to be the most influential. Subsequent examples often provide diminishing returns in focusing the model's output. 
Furthermore, given the input length limitations of language models, which restrict the number of exemplars in the prompt, 
we limit the number of references to 2.

\subsubsection{\textbf{Theme Description Generation}}
\label{sec:themedescriptiongeneration}
Given a user prompt \( In_p \), a UI layout \( In_l \), and the top \( k \) retrieved knowledge items \( \{ \sum_{i=0}^{k} refer_i\}\) (where, \( k=2 \)), these components are concatenated with the system's theme description prompt \( P_{theme} \) to formulate a comprehensive input for our agent. 


Upon completing the theme description generation, the resultant theme description is denoted by \( Res_{theme} \). 

\subsubsection{\textbf{Theme Image Generation}}
\label{sec:themeimagegeneration}
The objective of generating a theme image is to visually guide the overall design, allowing the sub-agents to generate coherent and consistent design.

We consider Stable Diffusion model~\cite{rombach2022stablediffusion}, a state-of-the-art text-to-image generation as our main model.
However, as this model only considers the text condition, it suffers from limited control over the spatial composition of the image, a crucial aspect for our UI design generation. 
To address this, we integrate ControlNet~\cite{controlnet}, which augments the diffusion model by providing enhanced spatial control over each module.
ControlNet controls the generation by manipulating the denoising module by importing additional spatial condition in UNet. 
We employ UI layout as the spatial condition. 


In addition, as the stable diffusion model 
faces obstacles when generating UI images, which requires a different domain knowledge from general images~\cite{chenObjectDetection}. We further finetune the model using the datasets of UI screenshots and their complementary high-level descriptions collected in Section~\ref{sec:knowledge}.

\subsubsection{\textbf{Sub-agent Execution}}

During sub-agent execution phase, Theme Design Agent identifies the optimal sub-agent corresponding to the component type. 
We considered 13 component types, as detailed by the Rico dataset. To elaborate, \( A_{text} \) handles ``Text Button'' and ``Text''. \( A_{img} \) is entrusted with ``Image'' and ``Background Image'', and the  \( A_{icon} \) focuses on ``Icon'' components. 
For other component types, we seek to render them editable, drawing insights from RaWi \cite{kolthoff2023data}. 
The color for any component is determined by identifying the dominant RGB color from the image region's histogram and then representing it in HTML code.

The dynamic between the central agent and the sub-agents is essential to our system's functionality. When a central agent engages a sub-agent, it uses a combination of past results and a cache pool to inform the sub-agent's prompts:

\vspace{-3mm}
\begin{equation}
Cache_t = Res_{t-1} + Cache_{t-1}
\end{equation}
\vspace{-8mm}

\begin{equation}
p_{t+1} = p_{sub} + Cache_t
\label{eq:subprompt}
\end{equation}
\vspace{-3mm}

In this context, \(Res_{t-1}\) is the output from the sub-agent for the \(t-1^{th}\) component. 
The \(Cache_t\) represents a cache pool that integrates the previous result with accumulated knowledge from earlier iterations. This cache pool serves as an essential memory function, retaining the design context and facilitating multi-turn interactions.
Meanwhile, \(p_{t+1}\) functions as the prompt for the \(t-1^{th}\) component's sub-agent interaction, incorporating both the specific prompt \(p_{sub}\) for the current sub-agent and the cumulative knowledge in \(Cache_t\). 


\subsection{Text Content Agent}
\label{textagent}
The primary role of \( A_{text} \) is to generate textual information tailored to specific GUI components. 
We use GPT-4~\cite{gpt4}.
The system prompt for this agent, represented as \(p_{text}\), is: \textit{``Based on the theme description and relevant details, provide a text content recommendation for the designated position at [bbox].''} 
In alignment with equation \ref{eq:subprompt}, the execution prompt of the \( A_{text} \) obtains its value from the central agent's cache, denoted as \(Cache_{t-1}\), and is subsequently concatenated with \(p_{text}\), to ensure consistency in system.

\subsection{Image Content Agent}
\label{imgagent}
To enhance the generation quality of local image-associated components and maintain the consistency, we also deploy our adaptively fine-tuned stable diffusion model in \( A_{img} \), but disabling ControlNet module.
we use the image description from the generated theme description as the user prompt (a).
Rather than using the latent image generated from Gaussian noise, we extract the area of the image component from the theme image as the input (b).
By harmonizing both textual and visual signals, we can guarantee that the produced content aligns seamlessly with the primary theme design intent.

\subsection{Icon Agent}
\label{iconagent}
\( A_{icon} \) is crucial for selecting appropriate icons and integrating them into the graphical user interface components.
In addition to acting as intuitive visual cues, well-designed icons can improve comprehension and the overall user experience. The system prompt for \(p_{icon}\), is: \textit{``In reference to relevant information and taking into account its positioning at [bbox], and based on the theme description, propose an indicative phrase like ``msg'' for the ``Icon''.}. 
As shown in equation (2), execution prompt is based on the central agent's cache, \(Cache_{t-1}\), combined with \(p_{icon}\). This approach ensures the icons selected match the GUI design semantically and visually. 



\section{Experiment}




\subsection{Experiments Setup}

From Section~\ref{sec:uiknowledge}, we in total collected a set of 3,738 UI textual descriptions, their corresponding wireframes and UI screenshots. 
For validation purpose, the UI screenshots are treated as the ground truth. It is essential to note that this data is reserved within the test set folder of Rico and has not been utilized during the model's fine-tuning phase. 
Images are resized to dimensions of $512 \times 512$.

To assess the quality and diversity of generated results, we utilize two metrics: Fréchet Inception Distance (FID)~\cite{2017GANs} and Generation Diversity (GD)~\cite{2019AIsketcher}, which are used in the image generation task~\cite{2019AIsketcher, 2020DoodlerGAN}. FID evaluates the similarity between generated results and real ones, while GD assesses the diversity of the generated prototypes.

\subsection{Results and Discussion}
\subsubsection{\textbf{RQ1: How does our fine-tuned model perform against baseline models in both quality and diversity?}}
\textit{ }

\textbf{Baselines.}
For RQ1, we consider two state-of-the-art image generation models: \textbf{stable-diffusion-1-5}~\cite{sd_v1_5} and \textbf{stable-diffusion-2-1}~\cite{sd_v2_1} 
as these two baselines do not incorporate ControlNet module, we consider variants of both by integrating ControlNet, denoted as \textbf{stable-diffusion-1-5 (with ControlNet)}, \textbf{stable-diffusion-2-1 (with ControlNet)}. 
In addition, we also employ an ablated version of our approach, \textbf{MAxProtytper (w/o ControlNet)} as a baseline.

\begin{table}[htbp]
\centering
 \caption{Comparative analysis of FID and GD scores among various models with and without the ControlNet}
  \label{tab:RQ1}
  \begin{tabular}{lcc}
    \toprule
    Model & FID & GD\\
    \midrule
     stable-diffusion-1-5 (w/o ControlNet) & 69.48 & 15.93 \\
     stable-diffusion-2-1 (w/o ControlNet) & 67.15 & 15.42 \\
     \textbf{MAxPrototyper (w/o ControlNet)} & \textbf{33.08} & \textbf{15.95} \\
    \midrule
     stable-diffusion-1-5 (with ControlNet) & 54.42 & 11.48 \\
     stable-diffusion-2-1 (with ControlNet) & 57.23 & 11.14 \\
     \textbf{MAxPrototyper} & \textbf{23.76} & \textbf{13.98} \\
  \bottomrule
\end{tabular}
\end{table}

\textbf{Results.}
As seen in Table \ref{tab:RQ1}, 
For generation quality, we consistently outperforms both baseline models in terms of FID scores, regardless of whether ControlNet is utilized. Specifically, a lower FID score suggests that the distribution of generated images more closely matches that of real images. This indicates that MAxPrototyper's outputs are more realistic, evident from its significantly reduced FID scores: 33.08 without ControlNet and an even lower 23.76 with ControlNet.

For diversity, Notably, GD scores reflect the model's ability to produce varied yet detailed UI designs. A higher GD indicates more detail, suggesting that the designs are diverse and intricate in their presentation. MAxPrototyper's GD scores, both with and without ControlNet, surpass those of the baseline models. This increase in GD values emphasizes our model's superior capability in producing designs that are varied and enriched with details compared to its peers.

For ControlNet,
Incorporating ControlNet results in noticeable improvements in FID scores for all models. For our MAxPrototyper, the FID shows an enhancement of 10.68, representing a substantial 32\% improvement. 

\textbf{In conclusion}, Our MAxPrototyper outperforms the baseline models in terms of both quality and diversity. The addition of ControlNet optimizes this performance further by introducing layout constraints, reinforcing its potential for generating realistic and detail-oriented UI designs.

\subsubsection{\textbf{RQ2: How do MAxPrototyper's individual modules influence its performance in quality and diversity?}}

\textit{}

\textbf{Baselines and Ablation Strategy.}
To better comprehend the interaction and individual impact of MAxPrototyper's components on the end result, we perform an ablation study, sequentially removing each module and evaluating the effect.
As seen in Table \ref{tab:RQ2}, we carefully crafted six ablations. 

\begin{table}
\centering
 \caption{Results of ablation study for different modules 
 }
  \label{tab:RQ2}
  \begin{tabular}{lcc}
    \toprule
     Method & FID & GD\\
    \midrule
    MAxPrototyper  & \textbf{23.76} & \textbf{13.98} \\
    \midrule
    -Retrieved Knowledge Items  & 42.56 & 12.14 \\
    -Theme Description Generation  & 28.43 & 11.77 \\
    -Theme Image Generation  & 33.08 & 12.95 \\
    -Text Content Agent  & 24.06 & 13.78 \\
    -Image Content Agent  & 24.71 & 13.38 \\
    -Icon Content Agent  & 24.32 & 13.41 \\
  \bottomrule
\end{tabular}
\end{table}
\vspace{-0mm}

\textbf{Results.}
MAxPrototyper's analysis highlights each module's critical role in achieving high-quality, diverse UI designs. The ``Knowledge Retrieval'' module is essential for data importation, with its removal increasing the FID score from 23.76 to 42.56, significantly impacting design quality. ``Theme Image Generation'' is vital for linking text and visuals, sees the FID score rise to 33.08 when omitted, underscoring its significance in visual integration. The ``Sub-agent Execution'' phase, with specialized agents for each design element, demonstrates the value of a detailed generation approach through minor FID score variations. 

\textbf{In conclusion}, ``Knowledge Retrieval'' and ``Theme Image Generation'' are the most influential, but all modules collaboratively enhance the final output.

\vspace{-4mm}
\section{Discussion and Future Work}

Our current work, MAxPrototyper, focuses on improving the design generation process and enhancing the quality of prototypes. It draws inspiration from a rich knowledge base and primarily relies on high-level user descriptions and wireframe layouts to translate user intent into tangible design prototypes. Looking ahead, there are several directions of improvements and expansions that we can do in the future:

\textbf{\textit{1) Automated Wireframe Generation.} }
To further streamline the user experience, we could offer automatic generation of relevant wireframes based on the user's high-level descriptions, thus removing one extra step from the user's side.
In the current iteration of our work, we chose not to include this feature, because we want to focus on refining the generation process of the design prototypes from provided wireframes and high-level descriptions. Integrating automated wireframe creation presents its own distinct challenges, particularly in ensuring that the created wireframes accurately and satisfactorily reflect the user's design intent.
However, recognizing its potential benefits, we regard it as a potential aspect for future enhancement.

\textbf{\textit{2) Dynamic Component Integration.}} 
Currently, our system mainly deals with static components, which we simply classify into text, images, and icons. Our future work can discuss about expanding this to include more dynamic and interactive elements, such as checkboxes and pickers, thus making our generated prototypes more functional and interactive.


\textbf{\textit{3) Tool Integration and Design-to-Code.}} 
With an aim to seamlessly incorporate our system into designers' workflow, we imagine MAxPrototyper could be developed as a plugin that integrates smoothly with popular wireframe sketching software, such as Figma~\cite{figma} and Sketch~\cite{sketch}. This would widen our system's functionality by providing designers with robust editing capabilities directly within these platforms. 
Additionally, future work could also explore ways to automate design-to-code conversion to provide designers and developers with a end-to-end solution.

\textbf{\textit{4) Usability and Accessibility.}} 
As we continue to improve MAxPrototyper, we are committed to augmenting our system with standard mobile device user interface guidelines like material design~\cite{materialdesign}. These improvements will further enhance our system's usability and accessibility, ensuring our prototypes not only look appealing but also offer intuitive and user-friendly interactions.

By focusing on these potential directions and continually refining our processes based on user feedback, we can better combine automation efficiency with creative freedom.

\bibliographystyle{ACM-Reference-Format}
\bibliography{reference}

\end{document}